\documentclass{aa}

\usepackage{graphicx}
\usepackage{natbib}
\bibpunct{(}{)}{;}{a}{}{,}
\usepackage{times}
\begin{document}
\title{Precision and accuracy of the orbital parameters derived from 2D \& 1D space observations of visual or astrometric binaries}
\titlerunning{2D \& 1D space astrometry of visual binaries}

\author{D.~Pourbaix\inst{1}\fnmsep\thanks{Post-doctoral Researcher, F.N.R.S., Belgium}}
\institute{Institut d'Astronomie et d'Astrophysique, Universit\'e Libre de Bruxelles, C.P.~226, Boulevard du Triomphe, B-1050 Bruxelles, Belgium}
\date{Received date; accepted date} 
\offprints{pourbaix@astro.ulb.ac.be}
\abstract{Recent investigations \citep[e.g][]{Han-2001:a} have shown that fitting the Hipparcos observations with an orbital model when the astrometric wobble caused by the companion is below the noise level can have rather unexpected consequences.  With new astrometric missions coming out within the next ten years, it is worth investigating the orbit reconstruction capabilities of such instruments at low signal-to-noise ratio.  This is especially important because some of them will have no input catalogue thus meaning that all the orbital parameters will have to be derived from scratch. The puzzling case of almost parabolic orbits is also investigated.
}

\maketitle

%
%------------------------------------------------------------------------------
\section{Introduction}
%------------------------------------------------------------------------------
%

Among the 23882 binaries of the Hipparcos Catalogue \citep{Hipparcos}, only 235 were processed with an orbital model (with seven additional parameters), the so-called DMSA/O entries.  However, the seven parameters were all fitted for only 45 of them.  In the other cases, the value of some parameters were assumed from previous investigations, mainly spectroscopic and interferometric orbits.  If the same proportion is assumed for GAIA, one will end up with about $5\,10^5$ systems for which one needs to derive an orbit from scratch.  However, simulations have shown that the number of systems observed by GAIA for which an orbit is worth deriving is about $10^{7}$ \citep{GAIA-RB}.

The significant improvement of the fit of DMSA/O observations when the orbital model is adopted does not mean they can constrain all the seven parameters.  Assuming the value of some parameters was not just useful, it was sometime necessary in order to derive a realistic orbit.  For instance, HIP~85749 is a well known spectroscopic binary \citep{Lucke-1982:a} with a period of 418 days.  Though almost three orbital revolutions took place during the Hipparcos mission and an orbital model is well appropriate, a 7-parameter $\chi^2$ fit leads to $e\approx 0.99$ whereas the radial velocities yield 0.21.

We report on the robustness of the fit of the seven parameters at different levels of noise on two dimensional and one dimensional (Hipparcos-like) observations.  From now on, it will be assumed that these observations have already been corrected for the parallactic effect and the proper motion, i.e. one will only deal with relative positions whose origin (center of mass or primary) is at rest.  Unlike \citet{Sozzetti-2001:a}, we limit ourselves to $S/N\sim 1$.

%
%------------------------------------------------------------------------------
\section{Synthetic material}\label{Sect:Synth}
\begin{figure}[thb]
\begin{center}
\resizebox{0.8\hsize}{!}{\includegraphics{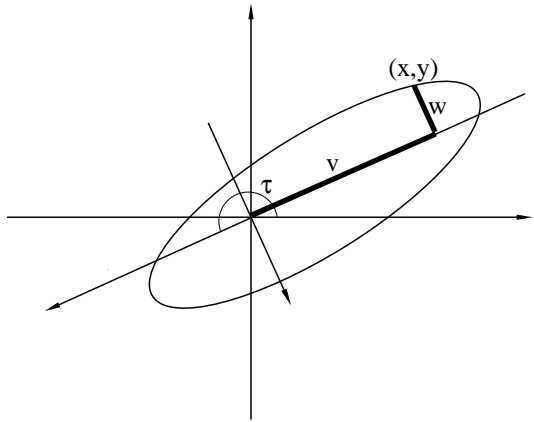}}
\caption[]{\label{Fig:descr}The scanning direction makes an angle of $\tau$ with respect to the $x$ axis.  $v$ is the signed projection along the scanning direction of the separation between the two components (or the component and the center of mass).  $w$ is the projection along the cross-scan direction.}
\end{center}
\end{figure}

%------------------------------------------------------------------------------
%
One thousand samples of $N_O$ observations ($N_O=60$ or 150) are generated by adding a Gaussian noise to the synthetic data of HIP~85749 (Table \ref{Tab:OrigOrbit}).  The photocentric orbit is assumed to correspond to the astrometric one ($a_0=a_1=a$).  The (groups of) observations (Fig.~\ref{Fig:descr}) are uniformly distributed over a period of $N_D$ days ($N_D=1000$ or 1600).  Two scanning laws are also investigated.  In one case, the orientation is uniform over $2\pi$ and the observations are uniformly distributed over $N_D$ days.  In the second, the system is observed on three consecutive revolutions of the instrument (the revolution period is set to 0.04 days), with the scanning direction slightly shifted (0.08 radian).  The groups of three observations are then uniformly distributed over $N_D$ days.  These properties, although realistic, do not specifically correspond to a particular forthcoming mission (DIVA, FAME, GAIA \& SIM).

The transformation of the Cartesian coordinates $(x,y)$ into the rectangular system attached to the instrument is:
\begin{eqnarray}\label{equ:defv}
v&=&x\cos\tau+y\sin\tau\\
w&=&-x\sin\tau+y\cos\tau
\end{eqnarray}
Several levels of noise ($\sigma_v=\sigma_w$) are investigated also: 0.25, 0.5, 1, 1.5, 2.0 and 2.5 mas.  Even if these values may sound very large with respect to modern instruments, they should be compared to the size of the projected orbit or $a$.  From now on, 2D and 1D solutions will refer to orbits derived from $(v,w)$ and $v$ only respectively.

\begin{table}[htb]
\caption[]{\label{Tab:OrigOrbit}Orbit used to generate the observations.  ABFG are the Thiele-Innes constants \citep{DoSt}}
\begin{tabular}{ll}\hline
Element  & Value \\ \hline
A (mas)  & +8.666885e-01\\
B (mas)  & -1.377152e+00\\
F (mas)  & +3.176942e+00\\
G (mas)  & -7.202854e-02\\
e        & 0.21\\
P (days) & 418.42\\
T (JD)   & 2443723.5\\
a (mas)  & 3.3264\\
\hline
\end{tabular}
\end{table}

For each pair of observations (1D and 2D), two least-square orbits are fitted using a local search only.  The starting point is always the orbit given in Table \ref{Tab:OrigOrbit}, i.e. the orbit used to generate the data.  This is an unrealistic situation where the true orbit is known and we expect it to be the solution of the local search procedure.  The distance between the solution of the local search and the true orbit is just a rough optimistic estimate of the reliability of the method.  Indeed, a global search could find a better solution (in terms of $\chi^2$) further away.  However, if one does already run into troubles with the local search, one can seriously start worrying for the future.

%
%------------------------------------------------------------------------------
\section{Orbit reconstruction}\label{Sect:Results}
%------------------------------------------------------------------------------
%

%
%------------------------------------------------------------------------------
\subsection{2D vs 1D data}\label{Sect:Results2D1D}
%------------------------------------------------------------------------------
%
\begin{figure}[thb]
\resizebox{\hsize}{!}{\includegraphics{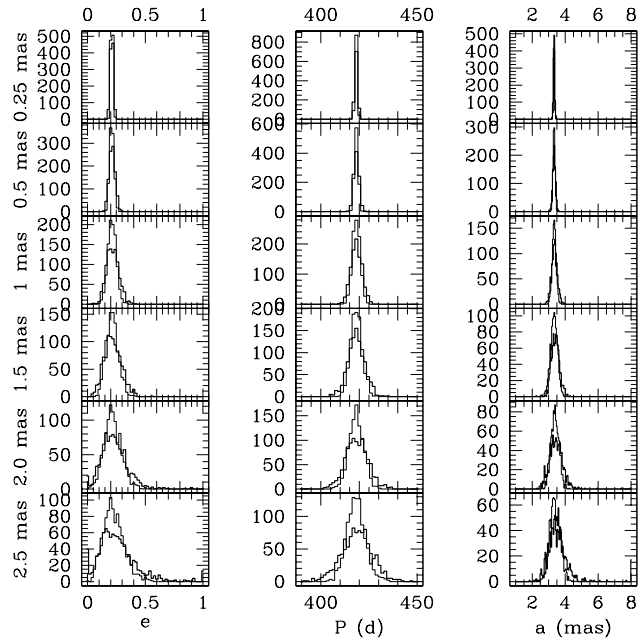}}
\caption[]{\label{Fig:dist}Distribution of the derived $e$, $P$ and $a$ from 2D (thin line) and 1D (thick line) data for different levels of noise with $N_D=1600$d, $N_O=150$ and the observations are grouped by 3}
\end{figure}

The distribution of $P$, $e$, and $a$ are given in Fig.~\ref{Fig:dist}.  The deterioration of the distribution from 2D to 1D and for decreasing $S/N$ is quite obvious.  For instance, the interval of periods with 1D at 1 mas is about as large as with 2D at 1.5 mas.  So, we do loose precision by only keeping the projection along the scanning direction.  However, as pointed out by \citet{Monet-1979:a}, the period remains rather well defined since it is independent of the Keplerian nature of the signal.

The behavior of the eccentricity is much worse.  Even if 1D and 2D distributions are still symmetrical at 1.5 mas, it is no longer the case at 2 mas and they start spreading all over the domain.  However, instead of becoming uniform over $[0,1[$, one notices an accumulation of circular and nearly parabolic orbits (lower left corner of Fig.~\ref{Fig:dist}).  As already mentioned, there is no way a global optimization method can improve such a behavior.

Even if $e$ gets close to 1, an adequate choice of $i$ and $\omega$ could nevertheless lead to a {\em reasonable} apparent orbit.  For instance, a very eccentric orbit with a large semi-major axis will look likely if $\omega$ and $i$ are close to ${\pi\over 2}$.  The lower right panel of Fig.~\ref{Fig:dist} exhibits such orbits with large semi-major axis.

The results obtained from the different combinations of $N_D$, $N_O$, scanning laws and noise levels are summarized in Table \ref{Tab:sim1a}.  As expected, the width of the distributions of $e$, $P$ and $a$ increases with the observation uncertainty.  Although the relation between the width and the uncertainty is rather smooth, almost linear, there is a noise threshold above which the distribution width, i.e. formally the parameter uncertainty, does explode.  In the smooth regime, the improvement caused by the second coordinate ranges from 40 to 60\%.  The effect of the scanning law is barely noticeable.

\begin{table*}[htb]
\caption[]{\label{Tab:sim1a}Dispersion of the derived $e$, $P$, \& $a$ with respect to the number of observations ($N_O$), the duration of the observation campaign ($N_D$ days) and the scanning law (SL).  SL=0 means that the $N_O$ points are uniformly distributed over $N_D$.  SL=1 corresponds to observations grouped by three, the groups being uniformly distributed overs $N_D$.  Regardless of SL, the scanning direction is uniformly distributed over $2\pi$.}
\setlength{\tabcolsep}{1.2mm}
\begin{tabular}{lll|lrrlrr|lrrlrr}\hline
&&& \multicolumn{6}{c}{$N_D=1000$d} & \multicolumn{6}{c}{$N_D=1600$d}\\
&&& \multicolumn{3}{c}{2D} & \multicolumn{3}{c}{1D} & \multicolumn{3}{c}{2D} & \multicolumn{3}{c}{1D} \\ 
$N_O$ & SL &  $\sigma$ &$\sigma_e$  & $\sigma_P$  & $\sigma_a$ &$\sigma_e$  & $\sigma_P$  & $\sigma_a$ &$\sigma_e$  & $\sigma_P$  & $\sigma_a$ &$\sigma_e$  & $\sigma_P$  & $\sigma_a$ \\
&& (mas) & & (d)  & (mas) &  & (d)  & (mas) & & (d)  & (mas) &  & (d)  & (mas)\\
\hline
60 & 0 &  0.25 & 0.01 & 1.31 & 0.05 & 0.02 & 1.86 & 0.07 & 0.01 & 0.75 & 0.05 & 0.02 & 1.12 & 0.07\\
60 & 0 &  0.50 & 0.03 & 2.65 & 0.10 & 0.05 & 3.91 & 0.15 & 0.03 & 1.50 & 0.10 & 0.04 & 2.22 & 0.15 \\
60 & 0 &  1.00 & 0.06 & 5.44 & 0.20 & 0.09 & 7.84 & 0.29 & 0.06 & 3.20 & 0.20 & 0.08 & 4.63 & 0.28 \\
60 & 0 &  1.50 & 0.09 & 7.84 & 0.33 & 0.15 & 12.01 & 364.70 & 0.09 & 4.68 & 0.30 & 0.14 & 7.32 & 344.91 \\
60 & 0 &  2.00 & 0.13 & 10.85 & 0.48 & 0.23 & 16.38 & 1266.08 & 0.12 & 6.30 & 0.44 & 0.25 & 10.22 & 1335.89 \\
60 & 0 &  2.50 & 0.20 & 14.29 & 777.02 & 0.31 & 25.39 & 2350.59 & 0.18 & 8.37 & 624.03 & 0.30 & 13.29 & 2101.33 \\
150 & 0 &  0.25 & 0.01 & 0.83 & 0.03 & 0.01 & 1.17 & 0.05 & 0.01 & 0.48 & 0.03 & 0.01 & 0.69 & 0.04 \\
150 & 0 &  0.50 & 0.02 & 1.62 & 0.07 & 0.03 & 2.37 & 0.09 & 0.02 & 0.95 & 0.06 & 0.03 & 1.36 & 0.08 \\
150 & 0 &  1.00 & 0.04 & 3.25 & 0.12 & 0.05 & 4.67 & 0.18 & 0.04 & 1.88 & 0.12 & 0.05 & 2.86 & 0.17 \\
150 & 0 &  1.50 & 0.06 & 5.15 & 0.19 & 0.09 & 7.11 & 0.29 & 0.06 & 3.01 & 0.18 & 0.08 & 4.21 & 0.25 \\
150 & 0 &  2.00 & 0.08 & 6.70 & 0.27 & 0.11 & 9.62 & 0.39 & 0.07 & 3.78 & 0.24 & 0.11 & 5.57 & 0.37 \\
150 & 0 &  2.50 & 0.10 & 8.29 & 0.33 & 0.17 & 12.02 & 650.23 & 0.10 & 4.79 & 0.33 & 0.16 & 7.48 & 809.25 \\
60 & 1 &  0.25 & 0.02 & 1.25 & 0.05 & 0.02 & 2.03 & 0.08 & 0.01 & 0.75 & 0.05 & 0.02 & 1.25 & 0.07 \\
60 & 1 &  0.50 & 0.03 & 2.58 & 0.10 & 0.05 & 4.21 & 0.17 & 0.03 & 1.50 & 0.10 & 0.05 & 2.48 & 0.15 \\
60 & 1 &  1.00 & 0.06 & 5.24 & 0.21 & 0.10 & 8.41 & 23.84 & 0.06 & 3.01 & 0.20 & 0.09 & 5.12 & 0.32 \\
60 & 1 &  1.50 & 0.09 & 7.62 & 0.31 & 0.17 & 13.22 & 536.81 & 0.09 & 4.77 & 0.30 & 0.18 & 8.86 & 361.40 \\
60 & 1 &  2.00 & 0.14 & 10.96 & 164.26 & 0.28 & 20.81 & 1443.31 & 0.13 & 6.67 & 4.20 & 0.25 & 11.75 & 1125.18 \\
60 & 1 &  2.50 & 0.18 & 13.54 & 615.85 & 0.32 & 26.13 & 2062.24 & 0.17 & 8.26 & 429.38 & 0.29 & 15.52 & 1766.75 \\
150 & 1 &  0.25 & 0.01 & 0.82 & 0.03 & 0.01 & 1.16 & 0.05 & 0.01 & 0.48 & 0.03 & 0.01 & 0.72 & 0.04 \\
150 & 1 &  0.50 & 0.02 & 1.66 & 0.06 & 0.03 & 2.40 & 0.09 & 0.02 & 0.92 & 0.06 & 0.03 & 1.36 & 0.09 \\
150 & 1 &  1.00 & 0.04 & 3.26 & 0.13 & 0.06 & 4.83 & 0.19 & 0.04 & 2.00 & 0.12 & 0.05 & 2.91 & 0.18 \\
150 & 1 &  1.50 & 0.06 & 4.95 & 0.19 & 0.09 & 7.35 & 0.28 & 0.05 & 2.99 & 0.19 & 0.07 & 4.42 & 0.27 \\
150 & 1 &  2.00 & 0.08 & 6.90 & 0.27 & 0.12 & 10.55 & 93.48 & 0.08 & 3.88 & 0.25 & 0.12 & 5.93 & 224.68 \\
150 & 1 &  2.50 & 0.10 & 8.36 & 0.34 & 0.18 & 13.19 & 743.73 & 0.09 & 4.91 & 0.31 & 0.17 & 8.00 & 732.08 \\
\hline
\end{tabular}
\end{table*}

The period covered by the observations and their number do affect the width of the distribution more strongly.  For instance, one remains in the smooth regime for all the 2D orbits with $N_O=150$.  As expected, the uncertainty on $P$ decreases with the length of the observation campaign.  However, the dispersion of the observations seems to play a role also, especially in the 1D case.  Thus, 150 observations uniformly spread over 1600 days give a better result, in terms of the region where the behavior of the distributions is smooth, than the same number of observations grouped by 3.  Actually, the latter case almost corresponds to 50 `normal' 2D points uniformly spread over the same period.

%
%------------------------------------------------------------------------------
\subsection{`Continuous' transition from 2D to 1D}\label{Sect:2D1D}
%------------------------------------------------------------------------------
%

So far, we have considered the ideal case of 2D observations with the same precision on the two axes in the one hand and the poor case of 1D data in the other.  We now consider 2D observations with different precisions along the two axes (along and cross scan).  Indeed, a slightly worse cross scan precision might still prevent from running into the same troubles as with 1D data at 2.5-mas noise level.

\begin{figure}[thb]
\resizebox{\hsize}{!}{\includegraphics{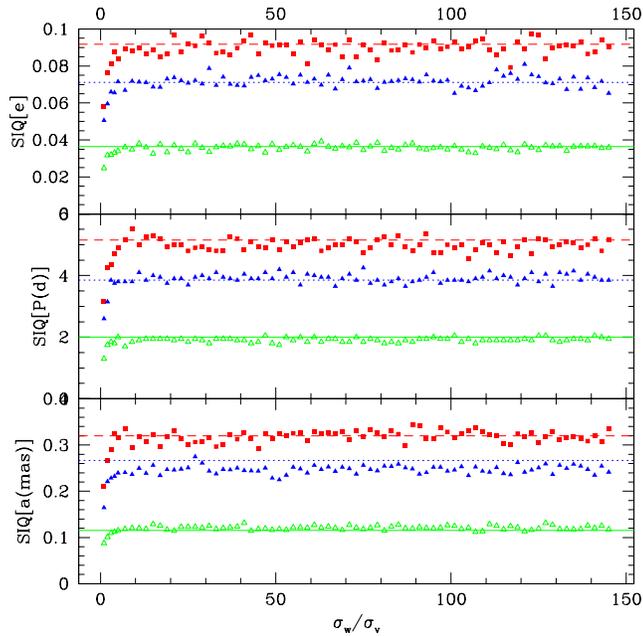}}
\caption[]{\label{Fig:evol2D1D}Evolution of the semi-interquartile of the distribution of $e$, $P$ and $a$ as a function of the cross-scan precision (reckoned in along-scan unit).  Filled squares, filled and open triangles correspond to $\sigma_v=2.5$, 2.0 and 1.0 mas respectively.  The long-dashed, short-dashed and continuous lines indicate the corresponding uncertainty with the 1D solution.}
\end{figure}

As seen in Table \ref{Tab:sim1a}, the sigma of the distribution of $P$ and $a$ can literally explode.  Moreover, $\sigma$ is not necessary a good indicator since there is no guarantee that the distribution is Gaussian, nor even symmetric.  In order to investigate that decrease of the cross-scan precision, we plot (Fig.~\ref{Fig:evol2D1D}) the semi-interquartile (a more robust indicator than $\sigma$) of the distribution versus the noise on $w$ (cross scan direction) reckoned in $\sigma_v$.  Only three different $\sigma_v$ are adopted since the point is just to see whether the overall behavior of $SIQ(\sigma_w/\sigma_v)$ is independent of the noise or not.

The first thing to notice is that the shape of $SIQ(\sigma_w/\sigma_v)$ is a flat line whose height corresponds to the 1D value.  Another important point is the very steep growth at small $\sigma_w/\sigma_v$.  The transition between 2D and 1D is thus very rapid.  The standard deviations of the distribution of $e$, $P$, and $a$ exhibit the same rapid change at small abscissae.

\begin{figure}[thb]
\resizebox{\hsize}{!}{\includegraphics{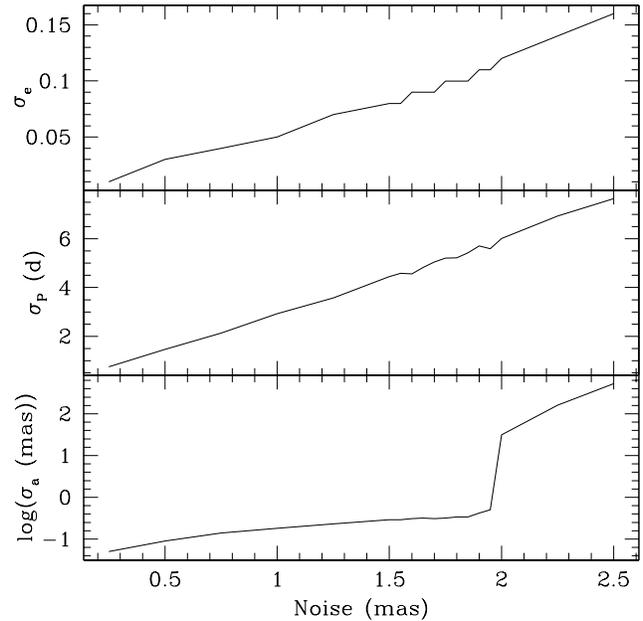}}
\caption[]{\label{Fig:evolnoise}Evolution of the standard deviation of the distribution of $e$, $P$, and $a$ at several noise levels}
\end{figure}

In Table \ref{Tab:sim1a}, we have shown that, regardless of the period duration, the number of observations, \dots\ there is a threshold on $\sigma$ above which the behavior of $\sigma_a$ explodes.  For Table~\ref{Tab:evolnoise} (and Fig.~\ref{Fig:evolnoise}), the investigation is limited to a 1600-day mission, with 150 observations grouped by three and a smaller increment for $\sigma$ between 1.5 and 2.0 mas is adopted in order to improve the identification of that threshold.  

The uncertainty on $e$ and $P$ looks like a linear function of the noise.  The same is true for $a$ up to about 1.9 mas ($S/N\sim 1.75$).  However, the ultimate goal of orbit reconstruction is to derive masses.  So, down to which $S/N$ can one go and still derive a useful result?  The relation between $a$ (actually $a_1$), the mass of the two components ($M_1$ and $M_2$), $P$ (in year) and the parallax $\varpi$ (reckoned in the same unit as $a$) is given by:
\begin{equation}
a=M_2(M_1+M_2)^{-2/3}\varpi P^{2/3}.
\end{equation}
Therefore, if $\varpi$ and $P$ are assumed to be error-free, $M_2(M_1+M_2)^{-2/3}$ has the same relative uncertainty as $a$.  Hence, although the fitted orbit still makes sense at $S/N\sim 1.9$, the theoreticians interested in masses should impose a larger $S/N$ in order to obtain useful results.

\begin{figure}[thb]
\resizebox{\hsize}{!}{\includegraphics{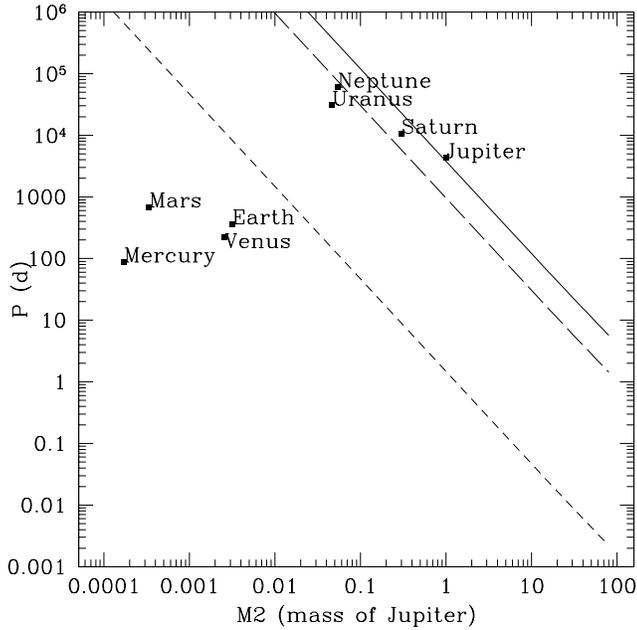}}
\caption[]{\label{Fig:M2P}Limits of orbit reconstruction for objects orbiting a solar-type star at 250~pc (solid line), 100~pc (long-dashed line) and 1.3~pc (short-dashed line) with $a=19\mu as$}
\end{figure}

What does this lead to with GAIA?  Like \citet{Lattanzi-2000:a}, we assume a noise level of $10\mu as$ for $V\le 12$ mag and a threshold at 1.9 for $S/N$.  We can thus expect to derive an orbit whose $a$ is as small as $19\mu as$.  The relation between $P$ and $M_2$ for objects with masses between $10^{-4}$ and 80 Jupiter masses, orbiting a solar-type star is plotted in Fig.~\ref{Fig:M2P}.  Our solar system is also represented.  Observed from $\alpha$ Cen, about 1.3~pc away, the four Jovian planets are detected but the periods are too long with respect to the mission duration.  In the other hand, the four inner most planets are still too light weight to cause a noticeable astrometric wobble.

\begin{table}[htb]
\caption[]{\label{Tab:evolnoise}Relative precision of $e$, $P$ and $a$ for some S/N levels (i.e. $a/\sigma$)}
\begin{tabular}{lllllll}\hline
Noise & $S/N$ & $\sigma_e$ & $\sigma_e/e$ & $\sigma_P$ & $\sigma_P/P$ & $\sigma_a/a$ \\
(mas) & &  & (\%) & (d) & (\%) & (\%) \\ \hline
0.25 & 13.32 &  0.01 &  4.8 &  0.75 &  0.2 &      1.5\\
0.50 &  6.66 &  0.03 & 14.3 &  1.46 &  0.3 &      2.7\\
0.75 &  4.44 &  0.04 & 19.0 &  2.13 &  0.5 &      4.2\\
1.00 &  3.33 &  0.05 & 23.8 &  2.93 &  0.7 &      5.4\\
1.25 &  2.66 &  0.07 & 33.3 &  3.58 &  0.9 &      6.9\\
1.50 &  2.22 &  0.08 & 38.1 &  4.45 &  1.1 &      8.7\\
1.55 &  2.15 &  0.08 & 38.1 &  4.59 &  1.1 &      8.7\\
1.60 &  2.08 &  0.09 & 42.9 &  4.56 &  1.1 &      9.3\\
1.65 &  2.02 &  0.09 & 42.9 &  4.82 &  1.2 &      9.6\\
1.70 &  1.96 &  0.09 & 42.9 &  5.05 &  1.2 &      9.3\\
1.75 &  1.90 &  0.10 & 47.6 &  5.21 &  1.2 &      9.6\\
1.80 &  1.85 &  0.10 & 47.6 &  5.22 &  1.2 &     10.2\\
1.85 &  1.80 &  0.10 & 47.6 &  5.42 &  1.3 &     10.2\\
1.90 &  1.75 &  0.11 & 52.4 &  5.71 &  1.4 &     12.6\\
1.95 &  1.71 &  0.11 & 52.4 &  5.59 &  1.3 &     15.0\\
2.00 &  1.67 &  0.12 & 57.1 &  6.01 &  1.4 &    944.7\\
2.25 &  1.48 &  0.14 & 66.7 &  6.94 &  1.7 &   4806.0\\
2.50 &  1.33 &  0.16 & 76.2 &  7.65 &  1.8 &  15963.4\\
\hline
\end{tabular}
\end{table}

%
%------------------------------------------------------------------------------
\section{Global vs local optimization}\label{Sect:GloLoc}
%------------------------------------------------------------------------------
%

Up to now, it was assumed that the actual orbit was available as an initial guess of the local minimization.  How good is this initial guess?  The simulations in Sect.~\ref{Sect:Results} show that the original orbit seldom minimizes the residuals (it is not the minimum of the least-square problem).  However, such simulations do not tell us whether that original solution is always in the vicinity of the lowest minimum or not.  Is the solution obtained with the local search from the original orbit the best one in terms of least-squares?  How reliable is the orbit with the lowest minimum?

The seven orbital parameters $(A, B, F, G, e, P, T)$ can be split into three groups: (a) those appearing in a linear way in the $\chi^2$ expression, namely $A, B, F$ \& $G$; (b) $P$ because it can be derived regardless of the the Keplerian nature of the problem \citep{Monet-1979:a}; (c) $e$ and $T$.  This distinction is therefore propagated to the way the $\chi^2$ is minimized.  The following scheme is thus adopted:
\begin{enumerate}
\item Guess $P$ using a period search technique \citep[e.g.][]{Horne-1986:a};
\item Explore the $(e,T)$-space and derive $A, B, F, G$ as the unique minimizer of $\chi^2$ when $e, P, T$ are fixed.
\end{enumerate}
For each trial of $e$ and $T$, another value of $\chi^2$ is evaluated.  In this approach, the dimension of the global optimization problem reduces to two.  That transformation speeds up the minimization process and increases the chance of getting the best minimum.  Several sophisticate methods exist for global search but, for the present purpose, a uniform 100 by 25 grid over $[0,1[\times[t_0,t_0+P[$ where $t_0$ is the time of the first observation seems to do just fine.  The orbit thus derived can then be used as the initial guess of a local search in the seven dimensional space of the parameters.

Once again, the method is assessed on synthetic observations.  However, contrary to what happened in Sect.~\ref{Sect:Results}, we limit ourselves on realistic cases that are likely to be difficult: an observation campaign of 1600 days, 150 observations grouped by 3 and a noise of $2.5$ mas.  1000 such sets of data are again generated.  For the sake of simplicity, we assume that the period search procedure returns the actual period (even if the $7p$ local search will update the period estimate later on).  The minima obtained from the true orbit in the one hand and from the grid search in the other were then derived.  The assumption about the efficiency of the period search technique might sound a bit optimistic and, indeed, it is.  However, as we are going to see, even if such an unexpectedly good estimate of $P$ is available, one can still run into troubles with other parameters.
  
The two minima are different in 22 cases only.  In two of them, the grid search gives a worse minimum than the one derived from the true orbit, thus indicating that a denser grid would give a better result.  However, with the present sparse grid, one already notices that 2\% of the solutions are not in the vicinity of the actual orbit.  A denser grid can only increase that percentage.

%
%------------------------------------------------------------------------------
\section{Almost parabolic orbits}\label{Sect:Parabolic}
%------------------------------------------------------------------------------
%

\begin{figure*}[thb]
\begin{center}
\resizebox{0.49\hsize}{!}{\includegraphics{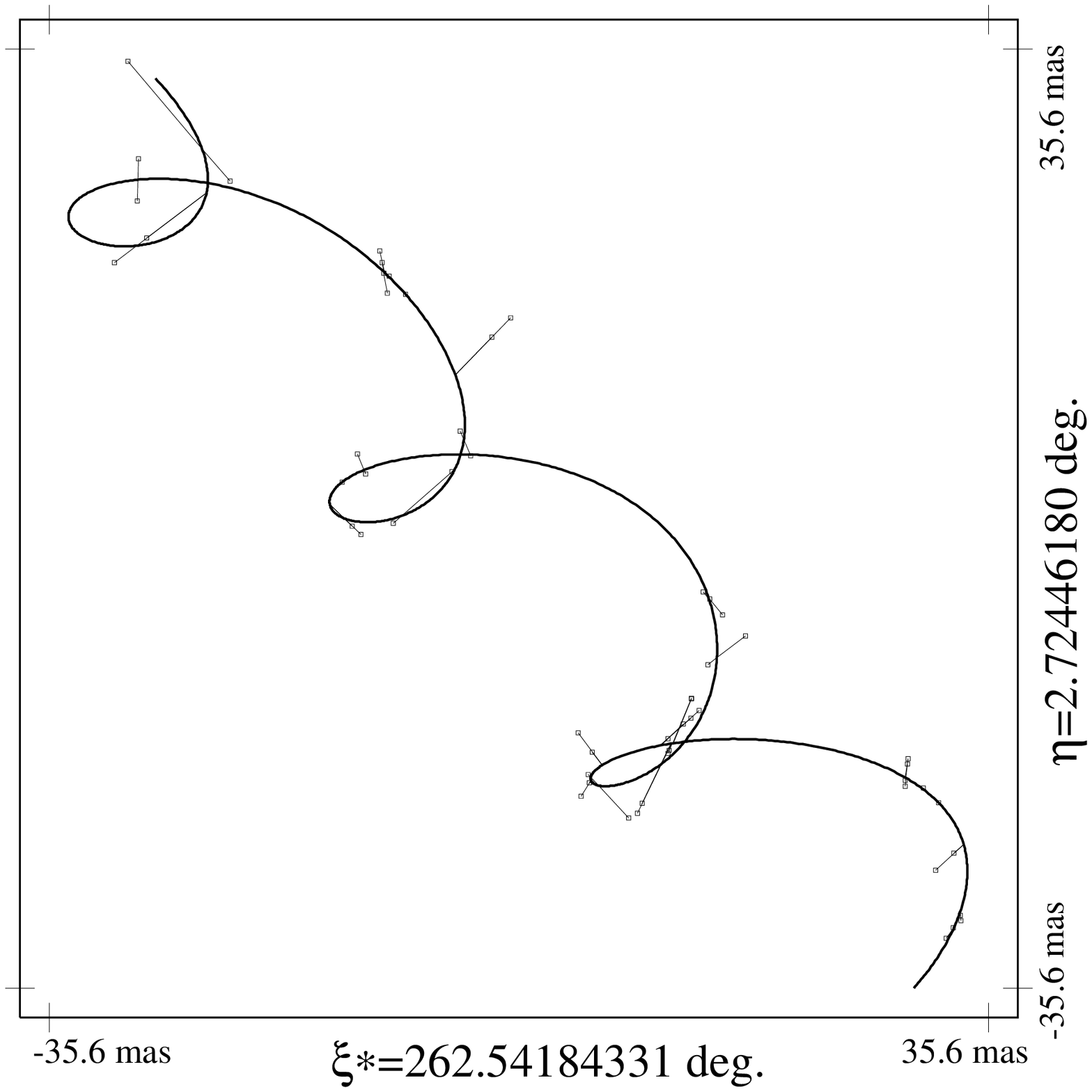}}
\resizebox{0.49\hsize}{!}{\includegraphics{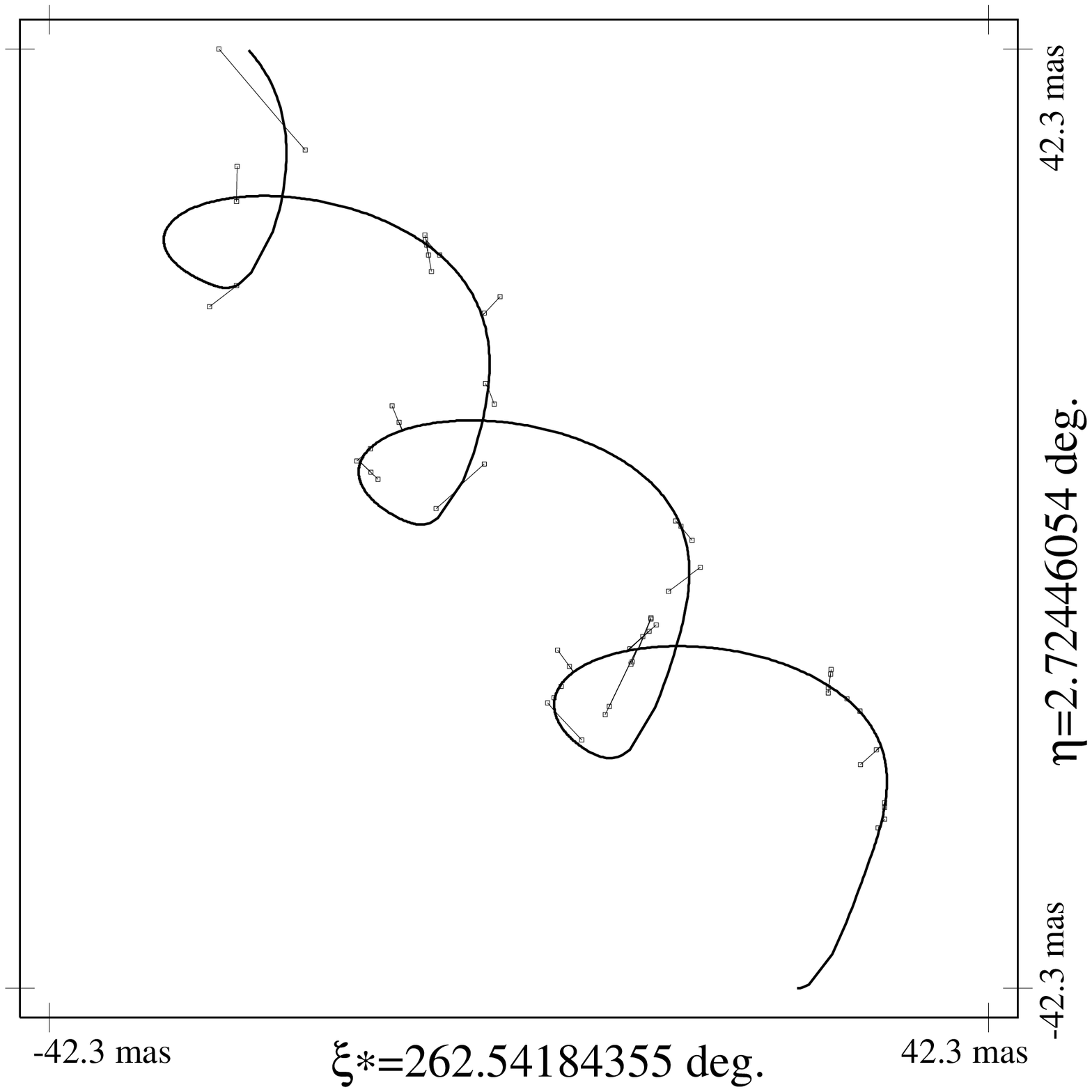}}
\caption[]{\label{Fig:parabola}An orbital model is worth adopting for HIP~85749 but rather similar apparent orbits nevertheless yield very different true orbits (the eccentricity is 0.2 and 0.998 for the left and right panels respectively).}
\end{center}
\end{figure*}

As already pointed out in Sect.~\ref{Sect:Results2D1D}, periodic though almost parabolic orbits (i.e. $e\approx 1$ and large $a$ both compensated by $i\approx\pi/2$) show up quite often even when the assumed or fitted orbital period does not exceed the mission duration.  When the seven orbital parameters are fitted, such orbits are quite usual and this is annoying and puzzling.  For instance, it is annoying because that would strongly bias the distribution of $e$ towards 1 if such solutions were published.  In the other hand, there seems to have no argument but statistical ones to discard these parabolic solutions.  Indeed, the apparent orbit looks realistic (the high inclination and eccentricity cancel out).  For instance, the two plots in Fig.~\ref{Fig:parabola} look similar although the eccentricity in the right panel is 0.998 where it is only 0.2 (spectroscopic value) in the left one.

Fitting the Hipparcos Intermediate Astrometric Data of the genuine single star HIP~115331 with a 12-parameter model (i.e. the five astrometric parameters plus the orbital ones) leads to $e=0.9999$.  However, in that case, the Thiele-Innes constants are not significantly departing from 0 \citep{Pourbaix-2001:b}.  Thus, even if one almost derives a parabola, the whole solution is discarded so it does not matter.  In the other hand, there are binaries for which the Hipparcos data leads to likely solutions (consistent with the spectroscopic orbits) even when the 12 parameters are adjusted from scratch (e.g., HIP~36377).

With Hipparcos, besides the single stars and the binaries where the whole set of orbital parameters makes sense, there are many spectroscopic binaries for which an orbital model is worth adopting but the eccentricity must be assumed otherwise it gets close to a parabolic solution.  Here, unlike the single star case, the Thiele-Innes constants are significantly non-zero.  Actually, when the seven parameters are fitted, $F$ and $G$ get much larger than when $e$, $P$ and $T$ are assumed.  In order to investigate the reasons of such a strange behavior, let us remind the equations of motion in the projected orbit \citep{DoSt}:
\begin{eqnarray}
x&=&A(\cos E-e)+F\sqrt{1-e^2}\sin E,\\
y&=&B(\cos E-e)+G\sqrt{1-e^2}\sin E
\end{eqnarray}
where $E$ is solution of
\begin{equation}
E-e\sin E=2\pi/P(t-T)
\end{equation}
From these equations, it is clear that when $e$ gets close to 1, $F$ and $G$ are no longer constrained.  Thus, $e\approx 1$ is a convenient way of getting rid of two parameters.  Is $e\approx 1$ a cause or a consequence of large $F$ and $G$?

The relation between the projected areal constant $\Gamma'$ and the areal constant in the true orbit $\Gamma$ is given by
\begin{eqnarray}
2\Gamma'\equiv\rho^2d\theta/dt&=&(AG-BF)\sqrt{1-e^2}2\pi/P \nonumber \\ 
 &=&a^2\cos i\sqrt{1-e^2}2\pi/P\\\label{Eq:Gamma} 
 &=&r^2 d\nu/dt\cos i \equiv2\Gamma \cos i \nonumber
\end{eqnarray}
where $\nu$ is the true anomaly in the true orbit.  It is actually because the law of areas holds in both the projected and true orbits that, in theory, the inclination can be derived.

With precise 2D observations of rather short period binaries, $\Gamma'$ is well constrained and so is $\cos i$ (the other elements being derived regardless of the law of areas).  When $S/N$ decreases, $\Gamma'$ gets less constrained.  If the observations are clustered, $\Gamma'$ is, on average, close to 0 which leads to a parabolic or edge-on orbit.  As already mentioned, the former has the advantage of getting rid of $F$ and $G$ in once.

With 1D data, the situation is worse.  Indeed, there is no way to estimate $\Gamma'$ and, therefore, $i$ cannot be derived unless:
\begin{itemize}
\item $e$ is set (for instance, thanks to a spectroscopic estimate)
\item or successive 1D data obtained with quite different scanning directions constrained 2D positions.
\end{itemize}
The latter case happens when, for instance, the period is long enough for the actual position of the companion to remain unchanged in between distant observations and short enough for the orbit to be well covered during the mission.  One way to achieve that would be to shorten the precession period of the satellite such that two consecutive observations would have quite distinct scanning directions.  Unfortunately, the precession period is no longer a free parameter in the design of the forthcoming missions.  Another favorable circumstance is when pairs of distant observations are separated by a multiple of the period.  This is rather unlikely but so seems to be the chance of fitting the Hipparcos data with a twelve parameter model from scratch.

In the case of GAIA, its spectroscopic capability should help constraining the eccentricity of, at least, the short period binaries.  Unfortunately, owing to the precision, this capability will be almost useless for the eccentricity of extra-solar planets.

%
%------------------------------------------------------------------------------
\section{Conclusions}\label{Sect:Conclusions}
%------------------------------------------------------------------------------
%

One cannot expect to do as much and as well with 1D observations as with 2D ones.  In between these two extremes, the second coordinate with a lower precision is not worth getting unless the ratio of the two precisions does not exceed 5.

The $\chi^2$ minimization can advantageously be replaced with a 3-stage evaluation thus yielding a global search in a 2-dimensional ($e$, $T$) space.  The period can indeed be independently guessed using period-search technique.  Owing to this low dimension, a grid approach is shown to be quite efficient even if one cannot prevent some false solutions to show up.

The results by \citet{Han-2001:a} have already warned the community about the reliability of the astrometric orbits, especially the inclinations, at low S/N even when some orbital parameters are adopted from spectroscopy \citep{Pourbaix-2001:a,Pourbaix-2001:b}.  The situation is likely to get worse when all the orbital parameters are derived from scratch.  All these improvements in the way the orbits are derived will remain useless unless one first finds a criterion for assessing the actual constraint on the projected areal constant and therefore on $i$ and $e$.

\begin{acknowledgements}
I thank Fr\'ed\'eric Arenou for his stimulating comments and for the fruitful discussions we have had so far about this topic.
\end{acknowledgements}

\end{document}